\documentstyle[12pt]{article}
\begin{document}
\pagestyle{plain}

\baselineskip=18pt

\begin{titlepage}

\vspace*{3cm}
\begin{center}
{\Large \bf Monte Carlo Studies of Ising Spin Glasses\\[1.5ex]}
{\Large \bf and Random Field Systems}\\[5ex]

{\bf Heiko Rieger}\\[1ex]

Institut f\"ur Theoretische Physik\\
Universit\"at zu K\"oln,\\
D-50923 K\"oln, Germany\\

\vskip 1.3cm
{\bf Abstract}\\[1.5ex]
\end{center}

We review recent numerical progress in the study of finite dimensional
strongly disordered magnetic systems like spin glasses and random
field systems. In particular we report in some details results for the
critical properties and the non-equilibrium dynamics of Ising spin
glasses. Furthermore we present an overview over recent investigations
on the random field Ising model and finally of quantum spin glasses.\\[4ex]

\begin{center}

{\bf To appear in:}\\
{\it Annual Reviews of Computational Physics II}\\
ed.\ D.\ Stauffer, World Scientific, Singapore (1995).

\end{center}

\end{titlepage}

\newcommand{\bc}{\begin{center}}
\newcommand{\ec}{\end{center}}
\newcommand{\ba}{\begin{array}}
\newcommand{\ea}{\end{array}}
\newcommand{\bt}{\begin{tabular}}
\newcommand{\et}{\end{tabular}}
\newcommand{\bi}{\begin{itemize}}
\newcommand{\ei}{\end{itemize}}
\newcommand{\be}{\begin{equation}}
\newcommand{\ee}{\end{equation}}

\section{Introduction}

Spin glasses and random field systems are magnetic materials in which
a structural disorder occurs as a consequence of a special
preparation process. The latter either changes the chemical
composition of compounds and alloys (via dilution or mixing) or it
decrystallizes the pure material (via sputtering or milling). The
effect is in any case a randomness in the position of the spins
and/or the sign and strength of the interactions among them. If this
disorder is strong enough it can drastically change the magnetic
properties of these material and can give rise to a new sort of phase
transitions. In many cases even a completely new kind of
low-temperature phase, the spin glass phase, might occur, which is in
some sense ordered but does not possess a translationally invariant
magnetic pattern. Theoretical models for these phenomena exist
for over 20 years, however, it took about half of this time to
establish the most fundamental facts for these models like the mere
existence of a phase transition. The characterization of the latter
transitions via the determination of the critical exponents is still
far from settled. The investigation of the low temperature phase and
its static and dynamic properties is a matter of ongoing experimental
and theoretical research. A major role in these activities is taken
by Monte Carlo simulations, which have found wide-spread applications in
condensed matter physics \cite{Binder92a}.

Due to the enormous research activity on spin glasses many excellent
reviews on this topic have already appeared during the last decade. To
begin with, Binder and Young \cite{Binder86} and Chowdhury
\cite{Chowdhury86} give a complete survey on the spin glass literature
until the year 1985.  Also the proceedings of the two Heidelberg
Colloquia on Spin Glasses \cite{Hemmen83,Hemmen86} contain useful
references. M\'ezard et al.\ \cite{Mezard87} focus on the mean-field
theory of spin glasses and the book by Fischer and Hertz
\cite{Fischer91} is a rather complete introduction into this complex
field. Mydosh's book \cite{Mydosh93} presents the most recent overview
over experiments on spin glasses and a very stimulating collection of
articles on random magnets \cite{Ryan92} gives insight into ongoing
experimental activities on spin glasses and random field systems.
Excellent reviews of the latter have been given by Nattermann and
Villain \cite{Nattermann88} focusing on theoretical aspects, by
Belanger and Young \cite{Belanger91} comparing theoretical predictions
for the critical properties with the experimental situation and by
Kleemann \cite{Kleemann93} about experimental evidence of the so
called domain states, which is a characteristic feature of these
systems.

Hence we do not intend to repeat the effort of the above reviews ---
we shall focus on frustrated Ising systems and its investigation via
Monte Carlo simulations during recent years.  In section 2 we review
the status quo concerning the critical behavior of finite dimensional
Ising spin glasses. Section 3 is devoted to the at present rapidly
evolving field of the non-equilibrium, low-temperature dynamics of
Ising spin glasses and other models exhibiting glassy features.
Section 4 surveys our present knowledge of the critical properties and
the off-equilibrium scenario of the random-field model and the diluted
antiferromagnet in a uniform field. In section 5 we review the
numerical methods that are useful in Monte Carlo studies of slowly
relaxing systems like those considered here. Section 6 discusses
quantum effects in spin glasses and especially the zero temperature
quantum phase transition occuring in transverse-field Ising spin
glasses.  Finally section 7 lists other models like Heisenberg,
Potts, orientational, vortex and Bose glasses.

\section{Ising spin glasses: Critical properties}

The simplest model that incorporates the necessary ingrediences for a
spin glass as quenched disorder and frustration is the Edwards-Anderson
model \cite{Edwards75}, which is defined by the Hamiltonian
(or energy function)
\be
H=-\sum_{\langle i,j\rangle} J_{ij} S_i S_j - h\sum_i S_i\;.
\label{ea}
\ee
The spin variables are of Ising type, i.e.\ $S_i=\pm1$, which are
placed on a $d$-dimensional hyper-cubic lattice (simple cubic in three
dimensions).  The interactions among them are short ranged, which is
indicated by the sum over only nearest neighbor pairs $\langle
i,j\rangle$.  The interaction strengths $J_{ij}$ are quenched random
variables that obey a given probability distribution, most commonly a
Gaussian with mean zero and variance one or a binary distribution
$J_{ij}=\pm1$ with probability one half. The last sum in (\ref{ea})
describes the effect of an external magnetic field if $h\ne0$.  A
typical experimental realization of this model is the short range
Ising spin glass Fe$_{0.5}$Mn$_{0.5}$TiO$_{3}$ \cite{Ito86,Gunnarson88}
mixing ferromagnetically and antiferromagnetically interacting components.

A complete survey on the results for the critical properties of the
two-, three- and four-dimensional Ising spin glass known until
1986 can be found in
\cite{Bhatt87}. In essence the spin glass transition (at temperature $T_c$)
in the infinite system is characterized by a divergence of the
non-linear (and not the linear, as in e.g.\ ferromagnets)
susceptibility
\be
\chi_{SG}=\frac{1}{N}\left[\sum_{ij}\langle S_i S_j\rangle^2\right]_{\rm av}
\sim(T-T_c)^{-\gamma}\;,
\label{susc}
\ee
which is caused by a divergence of the correlation length
$\xi\sim(T-T_c)^{-\nu}$ that sets the characteristic length scale for
the decay of spin correlations defined via
\be
G(r)=[\langle S_i S_{i+{\bf r}}\rangle^2]_{\rm av}
\sim r^{-(d-2+\eta)}\,\tilde{g}(r/\xi)\;.
\label{spatcorr}
\ee
Here $\langle\cdots\rangle$ means a thermal average and
$[\cdots]_{\rm av}$ indicates the average over the quenched disorder.

Two different routes were used to check numerically the above scenario
in three dimensions: One way is to perform Monte Carlo simulations of
finite but large systems above $T_c$, where the correlation length
$\xi$ is smaller than the system size $L$, and to perform a scaling
analysis of the spatial correlations $G(r)$. This has been done by
Ogielski and Morgenstern \cite{Ogielski85a} in extensive simulations
on a special purpose computer. The other way is to confine oneself to
smaller system sizes and to apply finite size scaling methods to
analyze the data for the spin glass susceptibility $\chi_{SG}$ and
certain combination of its moments, which has been done by Bhatt and
Young \cite{Bhatt85,Bhatt88}. Both approaches come to the conclusion that
there is indeed a finite temperature spin glass transition in the
three-dimensional Ising spin glass model (\ref{ea}) with a binary
distribution $J_{ij}=\pm1$, the critical temperature is
$T_c=1.175\pm0.025$, the correlation length exponent is
$\nu=1.3\pm0.1$ and the exponent for the spin glass susceptibility is
$\gamma=2.9\pm0.3$. The results from high-temperature series expansions
\cite{Singh86,Singh87,Wang88,Klein91}
agree with these values within the errorbars.

At a second order phase transition the characteristic time scale
$\tau$ for the decay of on-site correlations
\be
q(t)=[\langle S_i(t) S_i(0)\rangle]_{\rm }
\sim t^{-x}\tilde{q}(t/\tau)\quad{\rm with}\quad x=(d-2+\eta)/2z
\label{qt}
\ee
is also expected to diverge according to
$\tau\sim\xi^z\sim(T-T_c)^{-z\nu}$ with $z$ being the dynamical exponent.
One possibility of defining an effective relaxation time is
$\tau=(\int_0^\infty dt\,t\,q(t))/(\int_0^\infty dt\,q(t))$, for which
Ogielski \cite{Ogielski85b} finds $z=6.1\pm0.3$ for the 3d $\pm J$ spin
glass model. This is an unusually large value for $z$ (which causes
serious equilibration problems in Monte Carlo studies of the
critical properties of this model) and an exponential divergence
of $\tau$ at $T_c$ like $\tau\sim\exp\{A/(T-T_c)\}$ cannot be excluded.
Bernardi and Campbell \cite{Bernardi94a,Bernardi94b} perfomed a similar
analysis for other bond distributions and found differences for the
value of the exponent $x$ depending on the particular form of
these distributions, indicating that universality with respect to the
latter possibly does not hold. Hukushima and Nemoto \cite{Hukushima93}
investigated with the same method the two-dimensional case,
where an exponential divergence of $\tau$ at $T=0$ was found.

In a more recent work Blundell et al.\
\cite{Blundell92} perform a scaling analysis of the non-equilibrium
critical dynamics based on the hypothesis that the equal time
spin-spin correlation function measured at time $t$ after a
temperature quench from $T=\infty$ to $T=T_c$ obeys
\be
C(r,t)=[\langle S_i(t) S_{i+\bf r}(t)\rangle^2]_{\rm av}
=r^{-(d-2+\eta)}f(r/t^{1/z})\;.
\label{blund}
\ee
Note that here $\langle\cdots\rangle$ means an expectation value
with respect to the time dependent probability distribution generated
by the underlying stochastic process and not the thermodynamic equilibrium
expectation value. With other words, according to the hypothesis
(\ref{blund}) one does not need to equilibrate the system
to get estimates for the equilibrium critical exponents $\eta$ and $z$.
In this way a value $z=5.85\pm0.3$ and $\eta=-0.29\pm0.07$
was obtained, concurring with the value reported above by Ogielski
\cite{Ogielski85b}.

Another approach, suggested by Bhatt and Young \cite{Bhatt92}, is to look
at certain ratios of moments of time-dependent
spin autocorrelation functions as for instance
\be
R'(t)=\frac{[\langle N^{-1}\sum_i S_i(2t)S_i(t)\rangle]_{\rm av}}
{\sqrt{[\langle N^{-1}\sum_i S_i(2t)S_i(t)\rangle^2}]_{\rm av}}\;.
\ee
Since this is a dimensionless quantity it is expected to scale like
$R'(t)=\tilde{R}(t/\tau)$ with a characteristic time scale
$\tau=\tau(L,T)\propto L^z\tilde{\tau}\bigl(L^{1/\nu}(T-T_c)\bigr)$.
Again the determination of equilibrium critical exponents proceeds
via these hypothetical scaling forms of non-equilibrium quantities.
Assuming that $T_c=1.175$ (see above) in \cite{Bhatt92} it is found that
$z=6.0\pm0.5$, again in good agreement with Ogielski's result
\cite{Ogielski85b}.

As we have seen so far there seems to be rather compelling evidence
that the three-dimensional Ising spin glass in vanishing external field
has a second order phase transition at a finite temperature $T_c$.
However, a recent Monte Carlo study by Marinari et al.\
\cite{Marinari94} of the three-dimensional Ising spin glass on a
body-centered cubic lattice, where second and third nearest neighbor
interactions have been taken into account in order two increase the
hypothetical critical temperature, revealed a less clear picture: It
was found that in this model the data are compatible with a non-zero
temperature phase transition (with $T_c=3.25\pm0.05$, which is a
non-universal quantity and therefore different from the value reported
above, $\nu=1.20\pm0.04$ and $\gamma=2.43\pm0.05$) as well as with
a zero temperature transition:
\be
\ba{ccc}
\chi_{SG} & \sim & A [\exp(B/T)^4-1]+C\;,\\
      \xi & \sim & a [\exp(b/T)^4-1]+c\;.
\ea
\ee
Marinari et al.\ \cite{Marinari94} report that this form fits their
data better with a smaller systematic error, thus
giving new support for the speculation
that three dimensions may well be (instead of close to but larger than)
the lower critical dimension of the short ranged
Ising spin glass model.

In any other number of dimensions the situation is much clearer: In
two dimensions it was shown that no phase transition is present at any
finite temperature
\cite{Morgenstern79,McMillan83,Cheung83,Young83,McMillan84,Bray84,Huse85,Swendsen86,Bhatt88}.
For a $\pm1$ distribution the most recent results are from Bhatt and
Young \cite{Bhatt88} reporting that $T_c=0$, $\nu=2.6\pm0.4$
($\xi\sim T^{-\nu}$) and $\gamma=4.6\pm0.5$ ($\chi_{SG}\sim T^{-\gamma}$).
A continuous distribution yields significantly
different results indicating that this case is in a different
universality class than the $\pm1$ distribution, since the latter has
a large ground state degeneracy.  In four dimensions one finds much
stronger indications for the existence of a phase transition than in
three dimensions \cite{Bhatt88}: For a Gaussian distribution it is
found that $T_c=1.75\pm0.05$, $\nu=0.8\pm0.15$, $\gamma=1.8\pm0.4$
\cite{Bhatt88} and $z=4.8\pm0.4$ \cite{Bhatt92}. Furthermore the
extensive investigation of the probability distribution $P(q)$ of the
replica overlap $q$ \cite{Reger90,Parisi93c,Ciria93a}
revealed a picture that is
reminiscent of the Sherrington-Kirkpatrick (SK) model \cite{Sherrington75}
and its solution with broken replica symmetry
\cite{Mezard87}.
The upper critical dimension of the Ising spin glass, above which mean field
results should apply, is expected to be six \cite{Harris76}. This
prediction has been confirmed by Monte Carlo simulations of the
six-dimensional Ising spin glass with $\pm1$ couplings \cite{Wang93},
where at $T_c=3.03\pm0.01$ mean field critical behavior is found, which is
$\nu=1/2$ and $\gamma=1$, with logarithmic corrections due to being
{\it at} the upper critical dimension.

The effect of an external magnetic field on the critical
behavior of the Ising spin glass is also far from clarified.
The mean field scenario \cite{Mezard87} would suggest the existence of
an AT (de Almeida-Thouless \cite{Almeida78}) line in the $h$-$T$--phase
diagram and thus a spin glass transition even within a non-vanishing
external field. On the other hand phenomenological models
\cite{Fisher86b,Fisher88a}
dispute the relevance of mean-field results for finite
dimensional spin glasses and predict the absence of a phase transition
in a field. All numerical results obtained so far for four dimensions
\cite{Grannan91,Badoni93,Ciria93b,Ritort94}
seem to be compatible with the existence of a spin glass transition
at a nonvanishing temperature in a low enough external field.

In three dimensions the situation is still unclear: Kawashima et al.\
\cite{Kawashima92,Kawashima93} have investigated the variance of the
probability distribution of the central spin magnetization as well as
the probability distribution of the replica overlap and observe that
for increasing system size it scales to zero in non-vanishing external
fields. This implies the non-existence of a phase transition.
Following a proposition by Singh and Huse \cite{Singh91},
Grannan and Hetzel \cite{Grannan91} looked at the contour lines
$\chi_{SG}(T,h)=1$ in the $T$-$h$ diagram and
observe an upwards bend (i.e.\ $h\to\infty$)
for $T\to0$ in four dimensions (which indicates a spin glass phase)
and a downwards bend (i.e.\ $h\to0$) in two dimensions (meaning
the absence of a spin glass phase). This is depicted in figure
\ref{fig_cont}. The three-dimensional case
is only marginally bending upward, suggesting that three is close to
but larger than the lower critical dimension also in an external field
(see also \cite{Hetzel93} for the same study of the 3d Ising spin glass on the
face-centered cubic lattice).
Ritort \cite{Ritort94}  came to simlar conclusions by investigating
the quantities describing the sensitivity of the spin glass phase with
respect to magnetic-field perturbations.
Finally Caracciolo et al.\
\cite{Caracciolo90a,Caracciolo90b,Caracciolo92}
presented results of Monte Carlo
simulations of the (three-dimensional) system
\be
H=-\sum_{\langle i,j\rangle} J_{ij}(\sigma_i\sigma_j+\tau_i\tau_j)
+\sum_i h(\sigma_i+\tau_i)+\varepsilon\sum_i\sigma_i\tau_i\;,
\ee
which is identical to the two-replica method used by Bhatt and Young
\cite{Bhatt85} apart from the existence of a coupling of strength
$\varepsilon$ between the two replicated systems $\sigma$ and $\tau$.
This parameter $\varepsilon$ plays a similar role as the magnetic
field in a ferromagnet --- it is the conjugate field to the order parameter
$q$, the replica overlap. Hence one expects a discontinuity in
$q(\varepsilon)$ in the limit $\varepsilon\to0$ for temperatures
$T<T_c$, if there is a phase transition. The authors conclude that
their results are inconsistent with the predictions of the droplet
model \cite{Fisher86a} and support the mean field picture including
replica symmetry breaking and the existence of an AT-line (i.e.\ a
spin glass transition within a field) even in three dimensions.
These claims have been heavily disputed by Huse and Fisher
\cite{Huse91b} (see also the subsequent reply by Caracciolo et al.\
\cite{Caracciolo91}), the main counter-argument being that
their simulations are performed at temperatures {\it above} the
hypothetical AT-line and that the system sizes are much too small
to give reliable results for the low field regime. We will return to
the issue of the correctness of the droplet theory in finite dimensional
spin glasses below.

\section{Ising spin glasses: Non-equilibrium dynamics}

Most experiments on spin glasses at low temperatures are performed
in a non-equilibrium situation due to the astronomically large
equilibration times. Since the seminal work of Lundgren et al.\
\cite{Lundgren83} it has been realized by the experimentalists
\cite{Lundgren85,Hoogerbeets85} that magnetic properties of spin
glasses strongly depend on the time they spent below the spin glass
transition temperature $T_c$, a phenomenon that has been dubbed {\it aging}.
(see e.g.\ \cite{Ledermann91,Vincent92} for a review
about the experimental situation).
Figure \ref{fig_age} shows the result of a typical aging experiment:
The spin glass is rapidly cooled to low temperatures
in an external field $h$ (in three dimensional spin glasses below the
transition temperature) and the field is switched off only after
a macroscopic (several hours) waiting time $t_w$. As soon as
the field is zero (which defines the time $t=0$) the time-($t$)-dependence
of the thermoremanent magnetization $M_{\rm TRM}(t,t_w)$ is measured.
Usually (as for instance in simple ferromagnets) the equilibration
time is microscopic and for macroscopic waiting times one would
measure always the same magnetization curve $M_{\rm TRM}(t,t_w)$
independent of $t_w$. As can be seen in figure \ref{fig_age}
this is completely different in spin glasses, which is an
obvious manifestation of the huge time scales of the glassy dynamics.
In fact a spin glass transition is not a necessary
ingredient for this scenario to occur, it has also been
reported in experiments with two-dimensional spin glasses
like Ru$_2$Cu$_{0.89}$Co$_{0.11}$F$_4$ \cite{Dekker88,Dekker89}
or CuMn-films \cite{Mattson93} and in models without
frustration \cite{Rieger94d} or disorder \cite{Kisker94a,Cugliandolo94c}.
The observation characteristic for aging can also be made
in other amorphous or strongly disordered materials like
polymers \cite{Struik78}, charge-density-wave systems
\cite{Biljakovic89,Biljakovic91} or high-temperature superconductors
\cite{Rossel89}.

In Monte Carlo simulations the spin autocorrelation function
\be
C(t,t_w)=\frac{1}{N}\sum_{i=1}^N [\langle S_i(t+t_w)
S_i(t_w)\rangle]_{\rm av}
\label{ctw}
\ee
is the quantity that is most convenient to investigate the
non-equilibrium dynamics of spin systems. Here, as in equation
(\ref{blund}), $\langle\cdots\rangle$ formally means an average over
various thermal histories and initial conditions.  However, as long as
the number of samples for the disorder average $[\cdots]_{\rm av}$ is
large enough, it does not make a difference (numerically) if one
neglects this thermal average.

The limit $t_w\to0$, which corresponds
to experimental measurements of the remanent magnetization
$M_{\rm TRM}(t)$ after removing a strong
external field (of the order of the saturation field strength),
measures the configurational overlap with
a fully magnetized initial state.  Early investigators of this quantity
\cite{Binder76,Kinzel79,Kinzel86} observed already an algebraic
decay in a certain time window at low temperatures. This has
been confirmed recently for the SK-model \cite{Eissfeller92,Parisi93a}
and finite dimensional spin glass models
\cite{Huse89,Eissfeller92b,Parisi93a}.
In more extensive simulations for the 3d EA-model \cite{Rieger93b}
the temperature dependent exponent $\lambda(T)$ in
\be
M_{\rm TRM}(t)\sim t^{-\lambda(T)}\;,
\label{rem}
\ee
has been determined
for various temperatures below the spin glass transition temperature
and found a good agreement with those reported for the amorphous
metallic spin glass (Fe$_x$Ni$_{(1-x)}$)$_{75}$P$_{16}$B$_6$Al$_3$
\cite{Granberg87}, which is particularly well suited for measurements
of the remanent magnetization also for low temperature since its
saturation field strength is reasonably low (see figure \ref{fig_rem}).

For non-vanishing waiting times $t_w\ne0$ the function (\ref{ctw})
shows a behavior that is characteristic for aging phenomena.
Andersson et al.\ \cite{Andersson92}
and Rieger \cite{Rieger93b} showed that
qualitative features of experimental observations can be reproduced by
Monte Carlo simulations of finite dimensional spin glass models.
In addition to $C(t,t_w)$ the magnetic response
function $\chi(t,t_w)=M_{\rm TRM}(t,t_w)/h$ has been determined,
where $M_{\rm TRM}(t,t_w)$ is defined in the introductory paragraph
of this section.
Its logarithmic derivative $\partial\chi(t,t_w)/\partial\log\,t$
possesses a maximum at $t=t_w$, as observed in the corresponding
experiments \cite{Nordblad86,Granberg88}. Furthermore one observes that the
fluctuation--dissipation theorem
$\chi(t,t_w)=\beta\bigl(1-C(t,t_w)\bigr)$ is violated for
$t>t_w$. A more quantitative analysis of the the data for the function
$C(t,t_w)$, of which a typical set is depicted in figure \ref{fig_ctw},
reveals the following picture \cite{Rieger93b}:
In three dimensions there are strong indications for the scaling relation
\be
C(t,t_w)=t^{-x(T)}\Phi_T (t/t_w)\quad{\rm with}\quad
\Phi_T(y)\sim\left\{
\ba{ccc}
{\rm const.} & \;{\rm for}\; & y\to0\;,\\
y^{x(T)-\lambda(T)} & \;{\rm for}\; & y\to\infty
\ea
\right.
\label{scale}
\ee
to hold
This characteristic $t/t_w$-scaling has also been observed in
experiments with the insulating spin glass CdCr$_{1.7}$In$_{0.3}$S$_4$
\cite{Ocio85,Alba86}, in two-dimensional spin glasses
\cite{Rieger94b}, the SK-model \cite{Cugliandolo94a}, in simplified
spin glass models \cite{Bouchaud92,Parisi93b} and one-dimensional
models \cite{Rieger94d}. For a spin glass below the critical temperature
the relation (\ref{scale}) is expected to hold for any finite
waiting time $t_w$ (as long as the system is infinite), which expresses
the fact that for all temperatures below $T_c$ the equilibration time
is infinite --- in contrast to the situation in e.g.\ simple
ferromagnets. For $t_w\to\infty$ one obtains the
equilibrium autocorrelation function
$\lim_{t_w\to\infty}C(t,t_w)=q(t)\sim t^{-x(T)}$, see equation (\ref{qt}).
It should be noted that in spin glasses with replica symmetry breaking
a more complicated scenario can occur, see \cite{Franz94b} for a discussion.
At $T_c$ one has $x(T_c)\approx0.07$, which equals
the exponent reported for the equilibrium
dynamics of the 3d EA-model
\cite{Ogielski85b} as well as from experiments on the
short-range Ising spin glass Fe$_{0.5}$Mn$_{0.5}$TiO$_3$
\cite{Gunnarson88}. In other words: as long as $t\ll t_w$ one
observes a quasi-equilibrium dynamics (on time scales smaller
than the waiting time $t_w$) and the crossover at
$t=t_w$ signals the onset of the non-equilibrium behavior
characterized by an exponent $\lambda(T)$ that is significantly larger
than the corresponding equilibrium exponent $x(T_c)$ (e.g.\
$\lambda(T_c)\approx0.39$, for this value see also \cite{Huse89}).

At this point a short excursion on semantics might be
appropriate. {\bf A:} When speaking of equilibration time in the
preceeding paragraph we mean the equilibration time within one ergodic
component of an infinite system, like the one with positive (or
negative) expectation value for the magnetization in a ferromagnet
below $T_c$. This definition is independent of the answer to the
question, whether the phase space of a finite dimensional spin glass
splits into an infinite number of ergodic components separated by
infinite free-energy barriers below $T_c$ (as the SK-model) or not. We
observe that $\lim_{t\to\infty}C(t,t_w)\to0$ for any waiting time
$t_w$, but the more natural expectation for dynamics within an ergodic
component would be an exponential decay of $q(t)$ to a finite value
$q_{EA}$.  However, no indication for this has been reported yet
\cite{Ogielski85,Rieger93b,Cugliandolo94a}.

{\bf B:} The scenario (\ref{scale}) is what we call {\it aging}, and
we use the expression {\it interrupted aging} for a situation in which
we have some finite equilibration time $\tau_{\rm eq}$ in such a way
that (\ref{scale}) holds approximately as long as $t_w\ll\tau_{\rm
eq}$ and is simply replaced by true equilibrium dynamics
$C(t,t_w)\approx q(t)$ for $t_w\gg\tau_{\rm eq}$. This is what happens
in two-dimesnional spin glasses \cite{Schins93a,Schins93b,Rieger94b}
or random ferromagnets \cite{Rieger94d}, but also in pure systems like
the one considered in \cite{Kisker94a}, where depending on the system
parameters the time $\tau_{\rm eq}$ can be extremely large. Obviously
for an astronomically large $\tau_{\rm eq}$ neither experiments nor
Monte Carlo simulations might be able to discriminate between aging
and interrupted aging. It has also been noted \cite{Kisker94a} that in
the pure ferromagnetic Ising chain the aging scenario (\ref{scale})
holds at $T=0$ with $x=0$, thus aging phenomena do not rely upon
randomness or frustration althought the latter can greatly enhance it.
We would like to stick to this nomenclature in this review. However,
there exists also a different perception of what aging might be
\cite{Cugliandolo93,Cugliandolo94b,Mezard94,Franz94}, see also the
discussion following eq.\ (\ref{fdtratio}) below. In essence, aging
understood in this way is a property of off-equilibrium dynamics that
becomes manifest only in particularly chosen limiting procedure for
the times $t$ and $t_w$. A scenario like (\ref{scale}) with $x(T)>0$
would be called ``interrupted aging'', even if $\tau_{\rm eq}=\infty$.
This concept of aging is motivated by analytically tractable
mean-field models, where various infinite time limits can be done in a
straightforward manner --- its relevance for real spin glasses and
finite time experiments remains to be tested.

The underlying mechanism for the aging scenario (\ref{scale})
in finite dimensional
spin glasses is a slow domain growth, as suggested by Fisher and Huse
\cite{Fisher88b} proposing (ad hoc) a logarithmic growth law for the
characteristic domain size $R(t)\propto(\log t)^{1/\psi}$, and by
Koper and Hilhorst \cite{Koper88a}, stipulating (ad hoc) an algebraic
growth $R(t)\propto t^{\alpha(T)}$. Preliminary investigations on this
issue \cite{Huse91a} find some numerical support for the first
hypothesis, however, the above mentioned facts (the asymptotically
algebraic decay of $C(t,t_w)$ and the $t/t_w$ scaling) and recent very
extensive simulations pledge more in favor of an algebraic growth:
Rieger et al.\ \cite{Rieger94b} perform Monte Carlo
simulations of the two-dimensional EA-model with Gaussian couplings
and investigate the growth of spatial correlations via the
correlation function
\be
G(r,t_w)=\frac{1}{N}\sum_{i=1}^N
[\langle S_i(t_w) S_{i+\bf r}(t_w)\rangle^2]_{\rm av}\;,
\label{grt}
\ee
which becomes $G(r)$ of equation (\ref{spatcorr}) in the limit
$t_w\to\infty$. Note that the domain growth in various strongly
disordered systems like the site-diluted Ising model, the random field
Ising model and the random bond ferromagnetic Ising model, has been
investigated frequently via Monte Carlo simulations (see the review
by Chowdhury and Biswal in this series \cite{Chowdhury94}). These
models have the advantage that their ground state is known to be
ferromagnetic, which makes the identification of domains easy.
However, this is not the case for the present
system and such an investigation is much
more difficult here. Hence, a straightforward way to determine a
typical length scale for spatial correlations in a spin glass
is to calculate the function $G(r,t_w)$ defined in (\ref{grt}).

In order to take into account the square of the thermal average in
(\ref{grt}) in \cite{Rieger94b}
two replicas $a$ and $b$ of the system were simulated. Then
$[\langle S_i^a(t_w)S_i^b(t_w)S_{i+\bf r}^a(t_w)S_{i+\bf r}^b(t_w)
\rangle]_{\rm av}$
instead of $[\langle S_i(t_w) S_{i+\bf r}\rangle^2]_{\rm av}$ was
calculated, giving the same results with a much better statistics for
the first quantity. To improve the statistics one also has to average
over a suitably chosen time window around $t_w$. The correlation
length is defined via $\xi(t_w)=2\int_0^\infty dr G(r,t_w)$. It turns
out (see figure \ref{fig_alg}) that
\be
\xi(t_w)\propto t^{\alpha(T)}\;,
\label{alg}
\ee
with an exponent $\alpha(T)$ that decreases with temperature
yields a good fit to the time dependency of the correlation length.
Of course in the two-dimensional spin glass, which does not have
a phase transition at any finite temperature (see previous section),
$\xi(t_w)$ has to saturate at a finite value. This, however, happens
at small temperatures ($T\sim0.2$) only at very large times $t_w$
inaccessible to computer simulations. It should be noted that
a logarithmic fit $\xi(t_w)\sim(\log t)^{1/\psi}$ also works
fairly well --- with a temperature independent exponent
$\psi\approx0.63$. In recent experiments on the two-dimensional spin
glass Ru$_2$Cu$_{0.89}$Co$_{0.11}$F$_4$ \cite{Schins93a,Schins93b}
the scaling behavior of the ac-susceptibility has been analyzed
and it turned out that this is compatible with a prediction made by
Fisher and Huse \cite{Fisher88b} assuming a logarithmic growth
with $\psi\approx1.0$. Unfortunately no direct measurements of the
correlation length itself are available experimentally up to now.

For three dimensions the same kind of calculations as described above
have been done \cite{Kisker94b} (see also the work by Sibani and
Andersson \cite{Sibani94b,Andersson94b} discussed below) which
indicate the validity of algebraic growth law (\ref{alg}), too.  A
continuous set of exponents like $\alpha(T)$ for domain growth (or
sometimes called coarsening) might be disturbing especially if one
considers the work of Lai, Mazenko and Ma \cite{Lai88}. They classify
various non-equilibrium systems according to the underlying scaling
law for (free) energy barriers $B(L)$ that domains of size $L(t)$ have
to overcome in order to grow further: two of them lead to
$L(t)\propto\sqrt{t}$ and two lead to logarithmic growth --- the
fourth being identical to the proposal of Fisher and Huse
\cite{Fisher88b} for spin glasses $B(L)\sim L^\psi$, which
leads to $L(t)\sim(\log t)^{1/\psi}$ via activated dynamics.
However, as suggested by Rieger \cite{Rieger93b}, the scaling
law
\be
B(L)=\Delta(T)\log L
\label{log}
\ee
leads (via an activated dynamics
scenario in which it takes a time $\tau\sim\exp(B/T)$ to
overcome a (free) energy barrier $B$) to an algebraic domain
growth with a continuous set of exponents as in (\ref{alg}).
Moreover, following \cite{Fisher88b} in their argumentation,
we stipulate for the remanent magnetization that
$M_{\rm TRM}(t)\sim L(t)^{-\delta}$ and for the autocorrelation-function
that $C(t,t_w)\sim [L(t_w) / L(t+t_w)]^\delta$. Then it follows
$M_{\rm TRM}(t)\sim t^{-\lambda(T)}$,
$C(t,t_w)\sim t^{-\lambda(T)}$ for $t\gg t_w$ and also the $t/t_w$ scaling
stated in (\ref{scale}). The following table gives a collection of
predictions for the different scaling assumptions.
\vskip0.3cm

\hskip-0.4cm
\bt{|ll|c|cc|}
\hline
\rule[-3mm]{0cm}{0.9cm}
 & & Droplet-model \cite{Fisher88b} & MC-sim.\ & \\ \hline
\rule[-0mm]{0cm}{0.6cm}
Energy barrier & {\bf B} & $\Delta L^\psi$ & $\Delta(T)\log L$ & $\bullet$ \\
\rule[-0mm]{0cm}{0.6cm}
Activated dynamics & {\bf $\tau$}
                   & $\tau\sim\exp\,B/T$ & $\tau\sim\exp\,B/T$ &  \\
\rule[-0mm]{0cm}{0.6cm}
Domain size & {\bf R(t)} & $\left(\frac{T}{\Delta}\log\,t\right)^{1/\psi}$
                       & $t^{\alpha(T)}$ & ($\bullet$)  \\
\rule[-0mm]{0cm}{0.6cm}
Remanent Magnetization & {\bf M$_{TRM}$}
                     & $\left(\frac{T}{\Delta}\log\,t\right)^{-\delta/\psi} $
                       & $t^{-\lambda(T)}$ & $\bullet$  \\
\rule[-0mm]{0cm}{0.6cm}
Aging & {\bf C(t,t$_w$)}
      & $\overline{C}\left(\frac{\log t+t_w}{\log t_w}\right)$
      & $\tilde{C}\left(\frac{t}{t_w}\right)$ & $\bullet$  \\
\rule[-0mm]{0cm}{0.6cm}
Asymptotic decay & {\bf t$\gg$t$_w$} & $(\log\,t)^{-\delta/\psi}$
                                     & $t^{-\lambda(T)}$ & $\bullet$  \\
\rule[-3mm]{0cm}{0.9cm}
                 & {\bf t$\ll$t$_w$} & $(\log\,t)^{-\theta/\psi}$
                 & $t^{-x(T)}$  & ($\bullet$)  \\
\hline
\et
\vskip0.5cm

The bullets indicate that the Monte Carlo simulations described above
and below confirm the corresponding prediction in the last colomn.  A
bullet with brackets means that the numerical data can also be
interpreted according to the predictions of the droplet model.
Concluding the hypothesis (\ref{log}) seems to give a more consistent
description of experimental and numerical results on aging in spin
glasses presented so far.

Recently Sibani and Andersson \cite{Sibani94b,Andersson94b} gave
further support for the relation (\ref{log}) by means of the following
procedure: In Monte Carlo simulations of the two-- and three--dimensional
spin glass reference states $\Psi$ are generated with the help of a
careful annealing procedure down to $T=0$. Then a
certain amount of spins is flipped randomly under the constraint that
the energy difference of the new state to the old reference state
remains below a certain lid $b$. Via zero-temperature dynamics the
system has the opportunity to relax into a local energy minimum
configuration that is stable against single spin flips. Finally the
new state is analyzed by identifying all clusters of reversed spins.
For these clusters the size and energy distribution is recorded. It
turns out that the cluster volume increases exponentially (or even
superexponentially) with the lid $b$, implying a logarithmic dependence
for the energy barriers as in (\ref{log}). This dependency disagrees
with the result of Gawron et al.\ \cite{Gawron91}, who use an exact
search algorithm for calculating the energy barriers against inversion
of ground states. They obtain in two dimensions $B(L)\sim L$,
which means $\psi=1$. A serious
caveat in their approach might be the fact that their investigations
have been constrained to system sizes $L\le5$.

Especially with regards to the latter investigations it seems
that the systematic analysis of the phase-space structure of short
range spin glasses, especially their ground states and low lying
excitation, seems to become feasible with increasing computer power
These studies already gave valuable insights
\cite{Sibani93,Sibani94a,Klotz94a,Klotz94b},
supporting the phenomenological theories of hierarchical relaxation
for the spin glass dynamics \'a la Schreckenberg
\cite{Schreckenberg85}, Ogielski and Stein \cite{Ogielski85}, which
have been improved further by Sibani and Hoffmann
\cite{Sibani89,Schulze91,Hoffmann93} in order to model simple aging,
temperature step experiments and violation of the fluctuation
dissipation theorem. In particular with regards to improved
optimization algorithms
\cite{Groetschel86,Barahona94,Sutton94,Simone94}
(see also \cite{Kawashima92a} and the articles by Stariolo and Tsallis
\cite{Stariolo}, Hogg \cite{Hogg} and Tomassini \cite{Tomassini} in
this series) significant progress in the investigation of ground state
properties in spin glasses can be expected in the future.

Another procedure devised to test various phenomenological spin glass theories
are so-called temperature cycling experiments.  They consist of two
temperature changes during the time in which the material is aged in
the spin glass phase \cite{Refregier87}: either a short heat pulse is
applied to the spin glass during the waiting time after which the
relaxation of e.g.\ the thermo-remanent magnetization is measured, or
a short negative temperature cycle is performed, which is the same as
a heat pulse but with a negative temperature shift during the pulse.
It has been pointed out \cite{Lefloch92} that this kind of experiments
can discriminate between the droplet picture \cite{Fisher88b} and the
hierarchical picture \cite{Ledermann91} (see also \cite{Ioffe88}
and \cite{Freixa90} for a microscopic theorie of adiabatic
cooling, which are also capable to explain some of the obsevations).
The interpretation  of the experimental situation  is still controversial
(see also \cite{Weissmann93}
for a refreshing description of the present situation) --- essentially the
Saclay group pledging in favor of a hierarchical interpretation
\cite{Refregier87,Lefloch92,Vincent92} and the Uppsala group
taking the part of the droplet model \cite{Granberg90,Mattson93}. The
situation is the same with regards to numerical simulations of
temperature cycling experiments: Rieger \cite{Rieger94a} comes to
similar conclusions as the former and Andersson et al.\
\cite{Andersson94a} interpret their results in full agreement with
the latter experiments.  In both Monte Carlo simulations the same
quantities are studied (for instance the thermoremanent magnetization
$M_{\rm TRM}(t,t_{age})$, where $t_{age}$ now means the whole time in which
the temperature cycle has been performed), but the resulting data
are analyzed differently.
An alternative approach to the study of the temperature dependence of the
spin glass state has been suggested by Jan and Ray \cite{Jan94} using
damage spreading (for the use of the latter concept in Ising spin
glasses see \cite{Derrida87,Arcangelis89,Campbell90} and the article
by Jan and de Arcangelis in this series \cite{Jan94b}).

It might be difficult to make conclusive progress in this direction, because
these kind of experiments (real and numerical) always yield ambiguous results
that invite to one or the other interpretation.
However, the predictions of the droplet theory heavily
rely on the concept of {\it chaos} in spin glasses
\cite{Bray87,Nifle92,Nifle93}
(meaning the significant sensibility of the spin glass state to
infinitesimal changes of parameters like temperature or field) that
can be quantified by an overlap length $\zeta(T,\Delta T)$ via the
hypothesis
\be
[\langle S_i S_{i+r}\rangle_T
\langle S_i S_{i+r}\rangle_{T\pm\Delta T}]_{\rm av}
\sim\exp\{-r/\zeta(T,\pm\Delta T)\}\;,
\label{overlap}
\ee
where $\langle\cdots\rangle_T$ means the equilibrium expectation value
of one system at temperature $T$. The characteristic length scale
$\zeta(T,\pm\Delta T)$ should be finite for the positive and negative
sign of $\Delta T$ and it should decrease with increasing
$\Delta T$. There are quantitative predictions for these dependencies,
inaccessible to experiments, which can, however,
be calculated in Monte Carlo simulations. This seems to be a promising
endeavor. It should be noted that the existence of an overlap-length
as defined in (\ref{overlap}) is a prediction of phenomenological
concepts \cite{Bray87,Fisher88a,Fisher88b} {\it and} microscopic
theories on the basis of the SK-model, as has been reported by
Kondor \cite{Kondor89} and discussed further by Ritort
\cite{Ritort94}.

It should be mentioned that very recently very appealing
theoretical concepts for the non-equilibrium dynamics in spin glasses
have been developed. The phenomenological theory that fits the
experimental data for aging experiments in the best way up to now was
introduced by Bouchaud \cite{Bouchaud92,Bouchaud94a,Bouchaud94c}. In
essence it is a diffusion model in an abstract space in which each
state is characterized by a random (free) energy and hence by a
random, exponentially distributed trapping time. By arranging them in
a tree like structure, very reminiscent to the phase space structure that
emerges from Parisi's solution of the SK-model \cite{Mezard87}, one
obtains functional forms for $M_{TRM}(t,t_w)$ and $C(t,t_w)$ that
yield excellent fits to experimentally measured data.
Schreckenberg and Rieger \cite{Schreckenberg94} propose a different
diffusion model, which is based on an ultrametric tree that incorporates
a separation between quasi-equilibrium and non-equilibrium branches.
In this way aging occurs naturally via exploration of quasi-equilibrium
sub-branches of increasing depth.
Also the microscopic theory for the off-equilibrium dynamics of mean field
models of spin glasses has been pushed forward by Cugliandolo and
Kurchan \cite{Cugliandolo93,Cugliandolo94b} and Franz and M\'ezard
\cite{Mezard94,Franz94}: The mathematical difficulties for an
analytically exact solution of the dynamical off-equilibrium mean
field equations for e.g.\ the SK-model come from the lack of the
fluctuation-dissipation theorem (FDT) that relates autocorrelation and
response function (which allows the analytical solution in case of
equilibrium dynamics \cite{Sompolinsky81,Sommers87,Crisanti93,Kinzelbach93}).
The new approaches circumvent this by considering a so-called
fluctuation-dissipation ratio defined via
\be
x(t,t')= {r(t,t')\over \beta\partial C(t,t')/\partial t'}.
\label{fdtratio}
\ee
and postulating a particular set of properties for this function
$x(t,t')$ in various asymptotic limits, essentially setting up
an "ultrametric" for timescales. In this way Parisi's
static, equilibrium (!) order parameter function $q(x)$ finds
its counter-part in off-equilibrium dynamics.
It will be a challenging endeavor to test these
interesting ideas via Monte Carlo simulations and to check, whether
they might also be applicable in finite dimensions. The analysis of
Monte Carlo results for the three-dimensional EA spin glass
are compatible with the proposed scenario \cite{Franz94b} although,
due to the marginality of the three-dimensional case, further investigation
in higher dimensions are desirable and
would certainly give a much clearer picture.

Finally we would like to point out that aging and glassy dynamics have
gained much interest very recently. Apart from the issues
mentioned above in connection with spin glasses the central point of
the research activities is to model glassy behavior with spin systems
that have no quenched disorder (in order to try to understand the glass
transition that leads for instance to the formation of window glass):
among them are two- and three-dimensional models with competing nearest and
next-nearest neighbor interactions \cite{Shah90,Shore92}, the
anisotropic kagom\'e antiferromagnet \cite{Reimers93,Chandra93},
one-dimensional spin models with p-spin interactions
\cite{Kisker94a}, the Bernasoni model (also being a p-spin interaction
model, but with infinite range interactions)
\cite{Bouchaud94b,Marinari94b,Migliorni94,Krauth94a}
and certain field theoretial models with infinite range interactions
\cite{Cugliandolo94d,Franz94c}. Apart from the latter two, the bulk of
investigations of this very fascinating subject has to be done
numerically via Monte Carlo simulations.

\section{Numerical recipes}

It is obvious that thorough Monte Carlo simulations of spin glasses
require a huge amount of computer time. As long as one is interested
in purely dynamical phenomena as described in the previous section,
there is no other way than to implement some algorithm that
generates a stochastic process described by the Master equation
for the probability distribution of the spins. The possibility that
is most frequently used to achieve this,
is the heat bath, Glauber or Metropolis algorithm.
Here a random number generator is used to decide, whether a certain
spin configuration $\underline{S}$
is modified or not, according to predefined
transition probabilities, as for instance
\be
w(\underline{S}\to\underline{S}')={\rm min}\,
\{1,\exp(-\beta\Delta E)\}\;,
\label{prob}
\ee
where $\Delta E$ is the energy difference between the old
($\underline{S}$) and new ($\underline{S}'$)
configuration. For Ising systems very efficient implementations of
this algorithm with single spin flips have been presented that reach
a speed of $3\cdot10^8$ spin update attempts per second on a single
processor of a Cray YMP: for the pure Ising model by Ito and Kanada
\cite{Ito88} and Heuer \cite{Heuer90}, for the random field Ising model
\cite{Rieger93a} and finally for a whole class of Ising models
with or without quenched disorder defined via binary variables \cite{Ito93}
(see also the book of de Olivera \cite{Olivera91}
for a review on multi-spin coding techniques).
The enormous speed up compared to older multi-spin coding techniques
is based upon the use of a single random number for different (up to 64)
systems, like e.g.\ 64 different samples of random bond configurations
of the EA spin glass. This method does not work for a continuous
probability distribution of the quenched disorder variable (like
Gaussian) --- in this case more conventional methods have to
be used, which are at least one order of magnitude slower.
Although nearly all results reviewed above have been obtained with
these more or less optimized algorithms they reveal serious
deficiencies at low temperatures:

Equation (\ref{prob}) means that the new configuration
$\underline{S}'$ is rejected if a random number, equally distributed
between 0 and 1, is larger than $w(\underline{S}\to\underline{S}')$,
therefore this is called a rejection method. In system with a huge
number of local energy minimum configurations the number of rejections
becomes very large at low temperatures and most of the CPU time is
wasted by generating random numbers and calculating (or looking up in
a table) the transition rates. The only effect is to increment the
time that has passed since the beginning of the simulation by one
unit. Bortz, Kalos and Lebowitz \cite{Bortz75,Binder79} suggested already
20 years ago a method that seems to be more efficient: They proposed
to accept a new configuration {\it always} and then to increment
the time by a random number $\Delta t$ obeying a probability distribution
that is characterized by the sum over all probabilities for transitions
away from the old configuration:
\be
P(\Delta t)=\tau^{-1}\exp(-\Delta t/\tau)\quad
{\rm with}\quad
\tau^{-1}=\sum_{\underline{S}'} w(\underline{S}'\to\underline{S}')\;
\ee
Although this idea might be capable of circumventing the above
mentioned slowing down it has not been used very frequently up to now
\cite{Grest85,Shore92,Eissfeller92b,Krauth94b}. One of the problems
that typically occur also here is that once a transition away from a
local minimum configuration has taken place the system will very soon
return to it and thus decrementing the efficiency of the algorithm.
Very recently Krauth and Pulchery \cite{Krauth94a} proposed a variant
of this method that seems even to avoid this caveat of short cycles:
By keeping track of the configurations already visited (which means
storing them into the computer memory) the algorithm is forced to
generate new configurations in each iteration. The memory is cleared
as soon as a lower local energy minimum is encountered. In this way
they were able to explore times scales equivalent to $10^{13}$ conventional
Monte Carlo steps for a particular spin model with 400 spins.
This is very promising and would mean a real breakthrough if such
a performance could also be achieved with finite dimensional
spin glass models of reasonable size.

A completely different method that tries to avoid the tremendous
slowing down caused by coexisting energy minima in phase space with
large energy barriers between them is the so-called multicanonical
ensemble \cite{Berg92a} or simulated tempering \cite{Marinari92},
which is suited for the calculation of equilibrium
properties and also ground states of spin glasses.
Usually (as in the above mentioned algorithms) one performs an
importance sampling with the canonical
distribution $P_{\rm can.}(E)\sim\exp(-E/T)$, which is sharply
peaked around one (at high temperatures $T$), two (in case of first order
phase transitions) or several (in spin glasses or random field models)
values of the energy $E$. In order to escape from one energy minimum
region to explore others and, in particular for interface problems,
the regions in between them, one tries to generate an ensemble $P_{\rm
m}(E)$ that is approximately flat for the energy interval of interest,
especially between the sharp maxima of the canonical distribution.
This can be achieved with appropriately chosen functions
$\alpha(E)$ and $\beta(E)$ in the multicanonical ensemble
\be
P_{\rm m}(E)\sim\exp[-\beta(E)E+\alpha(E)]\;.
\ee
Since they are unknown a priori, they have to be determined
recursively during the simulations \cite{Berg92a,Berg92b}. Finally,
by re-weighting with $\exp[-E/T+\beta(E)E-\alpha(E)]$, the canonical
distribution can be reconstructed. With regards to spin glasses,
this ensemble has proven to be useful in particular for the investigation
of ground state properties \cite{Berg92b,Berg93,Berg94} (for this problem
see also the above mentioned works
\cite{Groetschel86,Barahona94,Sutton94,Simone94,Stariolo,Hogg,Tomassini}).

Cluster algorithms as suggested by Swendsen and Wang \cite{Swendsen87}
have become very useful in the investigation of pure systems (see
\cite{Swendsen92a} for an overview). However, they do not yet give a
significant improvement over local algorithms simulating single spin
flip dynamics in spin glass or random field models --- with at least
one exception: In a special implementation of a cluster algorithm for
the two-dimensional Gaussian EA-model Liang showed \cite{Liang92} that
the logarithm of the relaxation time is five times smaller than the
usual Metropolis algorithm. But he also pointed out that this
efficiency will not be reached in higher dimensions.  A cluster
algorithm for the random field Ising model has also been proposed
\cite{Dotsenko91}, but its efficiency remains to be tested.  The
reason for this is that due to randomness and frustration it is not
obvious at all how to construct spin clusters that can be reversed
with acceptable rates. Since the correct construction of these
clusters is the main problem in spin glasses we should also mention
the work on various cluster concepts in disordered systems
\cite{Stauffer93,Ray93,Stevens94,Gropengiesser94}.

Finally let us mention histogram techniques \cite{Ferrenberg89} that
also have been used frequently for pure systems \cite{Swendsen92a}: in
principle they allow to get information about thermodynamic quantities
of one system (i.e.\ one realization of the quenched disorder) in a
whole temperature interval by a Monte Carlo run at one single
temperature. However, this needs a lot of book keeping already for
pure systems, where several thousand configurations have to be stored.
In spin glasses and random field systems in addition one has to
perform an average also over this disorder, which might cause a
serious difficulty for data storage.

\section{Random field systems}

If one mixes a typical antiferromagnet like FeCl$_2$ with an
isostructural nonmagnetic material like CoCl$_2$ or NiCl$_2$ one
obtains diluted antiferromagnets, which are discussed in this series
\cite{Chowdhury94,Selke94}.  Within an uniform external field, however, a
large degree of frustration is induced and a completely new
universality class (with regards to critical properties) emerges.
Fishman and Aharony \cite{Fishman79} and also Cardy \cite{Cardy85}
pointed out that it is the same as the universality class
of the random field Ising model
(RFIM), which is defined by the Hamiltonian
\be
H= - J \sum_{\langle i,j\rangle} S_i S_j - \sum_i h_i S_i\;,
\label{rfim}
\ee
where the first sum is again over nearest neighbor pairs on
a d-dimensional lattice and the $h_i$ are independent
quenched random variables with $[h_i]_{\rm av}=0$ and $[h_i^2]_{\rm av}=h^2$.
A simple heuristic argument by Imry and Ma \cite{Imry75} shows that
the lower critical dimension of the RFIM (\ref{rfim}) should be $d=2$.
Indeed it has been proved rigorously \cite{Bricmont87} that
in three dimensions for small
enough random field strength $h$ there is ferromagnetic long range order
at low enough temperatures. Thus the existence of a phase transition
in three (and higher) dimensions is assured (in so far the situation
is slightly better than in spin glasses), however, there has been a
long lasting debate on the critical properties, which still awaits
a solution.

Villain \cite{Villain85} and Fisher \cite{Fisher86a} proposed a
scaling theory for the random field transition that relies upon
the assumption that random field induced fluctuations dominate over
thermal fluctuations at $T_c$. This implies for the singular
part of the free energy
\be
F_{\rm sing}\sim\xi^{\theta}\;,
\label{fsing}
\ee
where $\xi$ is the correlation length and $\theta$ is a new exponent.
Random field fluctuations alone produce typically an excess field of the order
of $\xi^{d/2}$ within a correlation volume, so a naive guess would be
$\theta$ roughly 1.5 in three dimensions. From (\ref{fsing}) the
modified hyperscaling relation follows
\be
2-\alpha=\nu(d-\theta)\;,
\label{hyper}
\ee
and it also implies an exponential divergence of the relaxation time
$\tau$ at $T_c$: $\tau\sim\exp(A/\xi^\theta)$. The decay
of the connected
($[\langle S_0 S_r\rangle]_{\rm av} -
[\langle S_0\rangle\langle S_r\rangle]_{\rm av}
\sim r^{-(d-2+\eta)}$) and disconnected
($[\langle S_0\rangle\langle S_r\rangle]_{\rm av}\sim
r^{-(d-4+\overline{\eta})}$) correlation functions at $T_c$
define two exponents $\eta$ and $\overline{\eta}$. These are
expected \cite{Villain85,Fisher86a}
to be related to the new exponent $\theta$ via
\be
\theta=2-\overline{\eta}+\eta\;.
\label{theta}
\ee
And, as usual, one has the scaling relations $\gamma=\nu(2-\eta)$,
$\beta=(d-4+\overline{\eta})$ and
$\alpha+2\beta+\gamma=2$. Obviously there seem to be three
independent critical exponents and a central issue of the activities
on the critical properties of the RFIM is the quest for an additional
scaling relation. Already Imry and Ma \cite{Imry75} conjectured that
$F_{\rm sing}\sim\chi$, where $\chi\sim (T-T_c)^{-\gamma}$ is the
susceptibility. This would imply $\theta=2-\eta$ (\ref{fsing}) and
therefore via (\ref{theta})
\be
\overline{\eta}=2\eta\;.\label{twoexp}
\ee
A set of analytical arguments by Schwartz et al.\
\cite{Schwartz85a,Schwartz85b,Schwartz86,Schwartz91}
supports this two-exponent scaling scenario indicated by (\ref{twoexp}).
Hence the exact Schwartz-Soffer inequality \cite{Schwartz85b}
$\overline{\eta}\le2\eta$ might be fulfilled as an equality.
Results of numerical studies are compatible with this picture:
In Monte Carlo simulations of the the diluted antiferromagnet
in a uniform field (DAFF) Ogielski and Huse \cite{Ogielski86a}
found $\eta=0.5\pm0.1$ and $\overline{\eta}=1.0\pm0.3$.
Ogielski \cite{Ogielski86b} investigated ground state properties
of the RFIM via combinatorial optimization methods \cite{Barahona82}
and found $\overline{\eta}=1.1\pm0.1$ and $\nu=1.0\pm0.1$.
Rieger and Young \cite{Rieger93} performed
the most extensive Monte Carlo simulation of the RFIM up to now
by sampling 1280 disorder configurations for each lattice size and
temperature and obtained via finite size scaling (for $h/T=0.35$)
\be
\ba{ccll}
\eta            & = & 0.56 & \pm0.03\\
\overline{\eta} & = & 1.00 & \pm0.06\\
\nu             & = & 1.6  & \pm0.3\\
\gamma          & = & 2.3  & \pm0.3\\
\beta           & = & 0.00 & \pm0.05\\
\alpha          & = &-1.0  & \pm0.3
\ea
\label{rfexp}
\ee
These are values for the binary $h_i=\pm h$ distribution of the random
fields, however, those obtained for a continuous (Gaussian)
distribution are not significantly different \cite{Rieger_u}.
In addition Dayan et al.\ \cite{Dayan93}
performed a real space renormalization group
calculation that gave identical values for $\eta$ and $\overline{\eta}$
as those listed in (\ref{rfexp}). Finally an extensive high-temperature
series expansion by Gofman et al.\ \cite{Gofman93} find
in three dimensions $\gamma=2.1\pm0.2$, concurring with the
Monte Carlo data (\ref{rfexp}) and
$\overline{\gamma}=\nu(4-\overline{\eta})=2\gamma$,
the last equality being a consequence of (\ref{twoexp}), which they
showed to hold also in 4 and 5 dimensions. Moreover, they demonstrated
an even stronger relation for the amplitude ratio to hold:
\be
A=\lim_{T\to T_c}\frac{\chi_{\rm dis}}{\chi^2(h/T)^2}=1\;,
\label{amplitude}
\ee
where $\chi_{\rm dis}=L^d[\langle S_i\rangle^2]_{\rm av}$ is the
disconnected susceptibility. The Monte Carlo data can be analyzed
in a similar way \cite{Rieger_u} and agree with (\ref{amplitude}).
Thus there are strong indications for
equation (\ref{twoexp}) to be correct and hence for the conjecture
of two exponent scaling to be true.

A closer inspection of the list of exponents (\ref{rfexp}) shows some
peculiarities: the order parameter exponent $\beta$ turns out to be
zero, which means that the magnetization jumps discontinuously to a
non vanishing value at the critical temperature. This hints at a first
order phase transition, a possibility that has already been suggested
by Young and Nauenberg \cite{Young85} (see however the interpretation
of X-ray scattering studies by Hill et al.\ \cite{Hill93}).  Note that
$\beta=0$ is also found for a continuous distribution \cite{Rieger_u},
which is important since in case of a binary distribution for larger
field strength a tricritical point is predicted in mean field theory
\cite{Aharony78}, which separates a first order (high fields) from a
second order (low fields) transition line in the
$h-T$-diagram. Usually, at a first order phase transition, one would
expect phase coexistence at $T_c$, which manifests itself in a typical
multi peak structure in the probability distribution $P(m)$ for the
order parameter \cite{Binder84,Challa86}. In the Monte Carlo data no
indication for such a scenario can be found \cite{Rieger_u}, instead
$P(|m|)$ shows a significant peak at a nonzero value for $|m|$ already
{\it at} the transition $T=T_c$ with $P(0)\to0$ in the thermodynamic
limit.

Furthermore, at a first order phase transition the specific heat
usually diverges with the volume of the system
\cite{Nauenberg74}, but the exponent $\alpha$ is {\it negative},
which means that the specific heat does {\it not} diverge at $T_c$.
By means of birefringence techniques Belanger et al.\
\cite{Belanger83,Belanger85,Pollack88} concluded from their experiments on
Fe$_{0.47}$Zn$_{0.53}$F$_{2}$ that $\alpha=0$ (i.e.\ a logarithmic
divergence of the specific heat). Only recently it was shown
\cite{Karszewiski94} that the same kind of experiments on
Fe$_{0.85}$Mg$_{0.15}$Br$_{2}$ are better compatible with a cusp like
singularity of the specific heat and $\alpha=-1$,
concurring with the value reported by in
\cite{Rieger93} see (\ref{rfexp}). However, this
value for $\alpha$ together with the other estimates in
(\ref{rfexp}) would violate the modified hyperscaling relation
(\ref{hyper}). Moreover, Schwartz \cite{Schwartz91b} derived an
exact inequality
\be
2-\alpha\le\nu(d-2+\eta)\;,
\ee
and accepting $\eta\approx0.5$ (since this result has a much smaller
errorbar than $\nu$) it would imply $\nu\ge2$, which is a rather large
value. Indeed in a recent Migdal-Kadanoff study Cao and Machta \cite{Cao93}
find such a large exponent $\nu=2.25$, and they also report $\alpha=-1.37$
and $\beta=0.02$, consistent with (\ref{rfexp}).
However, a value for $\alpha$ that is negative and large in modulus
causes serious difficulties with respect to the Rushbroke-equality
$\alpha+2\beta+\gamma=2$. Hence the random field enigma \cite{Shapir92}
is still far from being resolved.

In analogy to the spin glass research activity (see previous sections)
there has been considerable interest in the non-equilibrium dynamics
also in random field systems. It has been observed that diluted
antiferromagnets fall into a metastable domain state if cooled in
an external field $B$ below the critical temperature $T_c(B)$ (see the
article of Kleemann for a review \cite{Kleemann93}).  This domain
state has a finite correlation length that does not seem to increase
with time, in contrast to the continuous aging phenomena observed in
spin glasses (see section 3). Villain \cite{Villain84} and Grinstein
and Fernandez \cite{Grinstein84} have predicted a logarithmic growth
of the domain size (similar to what Fisher and Huse later predicted
for spin glasses \cite{Fisher88b}), and numerical investigations of
the domain growth in random field systems (e.g.\
\cite{Rao93,Puri93,Oguz94}) seem to be compatible with this prediction
(see also the review of Chowdhury and Biswal
\cite{Chowdhury94}). However, Nattermann and Vilfan
\cite{Nattermann88a} pointed out that DAFF map onto a RFIM {\it plus}
random bonds, and the latter produce an enormous pinning force so
that domain growth will only be observed on time scale up to $10^{11}$
years, which explains the experimental situation.  Moreover, these
time-persistent domains are fractal objects and have been analyzed in
Monte Carlo simulations of DAFF by Nowak and Usadel
\cite{Nowak91,Nowak92}.  Recently also hysteresis effects in the
RFIM at zero temperature have been studied \cite{Sethna93,Dahmen93}.

Nattermann and Vilfan also predict that after switching off the
external field in a DAFF the magnetization concentrated in the domain
walls will decay according to $M(t)\sim(\log t)^{-1/\Phi}$, which is
compatible with experiments \cite{Kushauer94a,Kushauer94b}. However
the results of Monte Carlo simulations fit better to an algebraic
decay \cite{Nowak89,Nowak91a} or an enhanced power law
$M(t)\sim M_0\exp\{-A(\log t/t')^y\}$ \cite{Han92a,Han92b}.

Very frequently it can be observed that by field decreasing
experiments with a DAFF a "stable" domain state already occurs for
field strengths $B$ larger than the value $B_c$ below which the system
orders antiferromagnetically (see \cite{Kleemann93} for an overview).
Thus in between the paramagnetic and the ordered phase a region for
this domain state has to be inserted in the corresponding phase
diagram (see figure \ref{fig_phase}).  This state has been identified
with an intermediate spin glass phase, characterized by dynamical
freezing and lack of long range antiferromagnetic order
\cite{Montenegro91,Belanger91b}.  From an
experimental point of view (as well as for Monte Carlo simulations
\cite{Nowak91}) it is still not clear, whether this intermediate
regime corresponds to an equilibrium phase.  However, only recently
M\'ezard and Young \cite{Mezard92} looked at the $N$-component version
of the RFIM in three dimensions and found in the limit $N\to\infty$
that replica symmetry is already broken {\it at} the ferromagnetic
transition at temperature $T_f$ (which is usually an indication for a
spin glass phase). In a subsequent investigation M\'ezard and Monasson
\cite{Mezard94b} were able to show that above $T_f$ a glassy phase
appears in a temperature interval $T_f < T <T_b$, $T_b$ being the
temperature above which the usual paramagnetic phase is entered.
Interestingly the ferromagnetic correlation length turns out to be
{\it finite}, reminiscent of a result by Guagnelli et al.
\cite{Guagnelli93}, who study numerically the solutions of the
mean field equations of the RFIM in three dimensions, and also very
similar to the experimental finding of finite domain sizes in the
intermediate domain state or spin glass region discussed above.
Thus the phase diagram of RFIM (or DAFF) seems to be much richer than
expected and in particular random field systems have much more in common
with spin glasses than anticipated up to now.

\section{Quantum spin glasses}

The models discussed so far are all describing classical systems
for which quantum fluctuations can be neglected. In most cases
this is correct, namely as long as $T_c>0$ since critical fluctuations
at the transition occur at a frequency $\hbar\omega\ll k_B T$
that is proportional to the inverse relaxation time $\tau$ and thus
approaches zero for $T\to T_c$ due to critical slowing down. Hence
any finite temperature will destroy quantum coherence and the
system will behave calassically. Very recently however, spin
glasses began to enter the quantum regime \cite{Sachdev94}.

The interesting theoretical question is: What are the effects of
quantum mechanics on the physics of strongly disordered systems at
zero temperature, where no heat bath is present and hopping over
energy barriers is replaced by tunneling them quantum-mechanically.
The best known and most studied example in this respect is the
zero temperature metal-insulator transition \cite{Lee85}. The renewed
interest in spin glasses in the quantum regime was kindled by a series
of recent experiments \cite{Wu91,Wu93} on the dipolar Ising magnet
Li$_{x}$Ho$_{1-x}$YF$_{4}$, where $T_c$ was driven down to zero by the
application of a transverse magnetic field $\Gamma$ (see the phase
diagram depicted in figure \ref{fig_qsg}). Experimental realizations of
quantum spin glasses are already known for more than 10 years: the so
called proton glasses \cite{Courtens84,Pirc87}, which are random
mixtures of ferroelectric and antiferroelectric materials like
Rb$_{1-x}$(NH$_{4}$)$_{x}$H$_{2}$PO$_{4}$. Here the proton position,
describable by an Ising spin variable, tunnels between two energy
minima with a fixed frequency modelled by a transverse field acting on the
spins. In the transverse field Ising magnets
mentioned above, however, it became possible to study the {\it zero}
temperature phase transition occuring for critical transverse field
strength $\Gamma_c$ by simply tuning the external field strength.
This quantum phase transition lies within a different
universality class than the usually studied (classical) spin glass
transition at finite temperatures and it turns out that their
properties differ significantly.

The above mentioned experiments can be described by the model
Hamiltonian of an Ising spin glass in a transverse field \cite{Wu91,Wu93}
\be
H=-\sum_{\langle ij\rangle}J_{ij}\sigma_i^z\sigma_j^z-
\Gamma\sum_i\sigma_i^x
\left(-h\sum_i \sigma_i^z\right)\;,
\label{tisg}
\ee
where the $\sigma_i$ are Pauli spin matrices, $\Gamma$ is the
strength of the transverse field and $h$ is a longitudinal
magnetic field used to define magnetic susceptibilities but usually
set to zero. Otherwise this model is
equivalent to the EA-model in $d$ dimensions defined in (\ref{ea}).
Obviously, for $\Gamma=0$ the quantum-mechanical Hamiltonian
(\ref{tisg}) is diagonal in the $z$-representaion of the
spin operators, which in this case can simply be replaced by their
eigenvalues $\pm1$ (after rescaling the couplings) giving exactly the
classical EA-model (\ref{ea}). In this way the transverse field
introduces the quantum mechanics into the spin glass problem and the
value of $\Gamma$ tunes the strength of the quantum fluctuations.  At
zero temperature and $\Gamma=0$ the system described by (\ref{tisg})
will be in its uniquely determined ground state, which is identical to
the classical ground state of (\ref{ea}). In this case one has
$\langle\sigma_i^z\rangle=\pm1$ for all sites $i$ and therefore
$q_{EA}=[\langle\sigma_i^z\rangle^2]_{\rm av}=1$ , where
$\langle\cdots\rangle$ means the quantum-mechanical expectation
value.

If we switch on the transverse field ($\Gamma>0$) the
Hamiltonian (\ref{tisg}) is not diagonal in the $z$-representation
any more and its ground state will be a superposition of the classical
ground state plus various excited states, which describes
the quantum-mechanical tunneling at zero temperature between the
local energy minima of the classical Hamiltonian. Furthermore
$|\langle\sigma_i^z\rangle|<1$ since the transverse
field tries to align the spins in the $x$-direction
and therefore $q_{EA}=[\langle\sigma_i^z\rangle^2]_{\rm av}<1$.
Increasing $\Gamma$ diminishes the EA-order parameter
$q_{EA}$ and for some critical value $\Gamma_c$ it will be
zero: $q_{EA}=0$ for $\Gamma\ge\Gamma_c$. This is the zero
temperature phase transition we are interested in and obviously we
cannot expect that its critical properties have anything in common
with the finite temperature classical spin glass transition discussed in
section 2.

In order to describe this zero-temperature transition one introduces a
quantity measuring the distance from the critical transverse field
strength (at $T=0$) $\delta=(\Gamma-\Gamma_c)/\Gamma_c$. If one
assumes a conventional second order phase transition one has
$q_{EA}\sim|\delta|^\beta$ and $\chi_{SG}\sim|\delta|^{-\gamma}$ for
$\delta\to0$. Spatial correlations decay on a characteristic length
scale that diverges at the critical point as usual:
$\xi\sim|\delta|^{-\nu}$ and these exponents defined so far would be
sufficient to describe the static critical behavior of a classical
spin glass transition.  However, at a zero temperature transition
driven solely by quantum fluctuations static and dynamic quantities
are linked in such way that the introduction of a characteristic
time-scale (or inverse frequency) is necessary (see e.g.\
\cite{Fisher89}): \be \xi_\tau\sim\xi^z\sim|\delta|^{-z\nu}\;,
\label{xitau}
\ee
where $z$ is the dynamical exponent. This will become more evident
below, when we consider an equivalent classical model.

Much work in the past has been devoted to the infinite range model
\cite{Bray80,Ishii85,Kopec88,Thirumalai89,Buttner90,Goldschmidt90,Usadel91}
and the phase-diagram in the $\Gamma$-$T$-plane, which looks similar to
the one shown in figure \ref{fig_qsg}: for low enough temperature $T$
and field $\Gamma$ one finds a transition line separating a
paramagnetic phase from a spin glass phase. Recently Miller and Huse
\cite{Miller93} and Ye, Sachdev and Read \cite{Ye93} focused on the
zero-temperature critical behavior and calculated the critical
exponents $\gamma=1/2$ (with multiplicative logarithmic corrections),
$\beta=1$ and $z\nu=1/2$. Interestingly it seems that this quantum
SK-model seems to fall into the same universality class as an
infinite range metallic spin glass model incorporating itinerant
electrons, which was investigated recently by Oppermann
\cite{Oppermann94}. Another proposition for a mean-field quantum spin
glass that is exactly solvable was made by Nieuwenhuizen
\cite{Nieuwenhuizen94} via the introduction of a quantum description
of spherical spins.

In one dimension the situation is quite different. The transverse
Ising chain with random bonds and/or random transverse fields can be
mapped (see below) onto the McCoy-Wu model
\cite{McCoy68,McCoy69a,McCoy69b}, for which various analytical results
are known
\cite{Hoever81,Shankar87,Nieuwenhuizen89,Fisher92,Mikheev94,Mikheev94b}.
In particular D.\ Fisher \cite{Fisher92} has shown within a
renormalization group calculation that {\it typical} and {\it average}
spatial correlations behave differently: the typical correlation
length diverges with an exponent $\tilde{\nu}=1$ (as also found by
Shankar and Murthy \cite{Shankar87}), whereas the average correlation
length diverges with an exponent $\nu=2$, thus obeying the rigorous
inequality $2/\nu\le d=1$ \cite{Chayes86} as an equality.  Due to
activated dynamics it turns out that $\xi_\tau\sim\exp(A\sqrt{\xi})$
and therfore $z=\infty$. Furthermore it is predicted that
$\beta=(3-\sqrt{5})/2\approx0.38$ and that, similar to the
surface magnetization \cite{McCoy69b}, the bulk magnetization behaves
nonanalytically as a function of the field $h$ already above the
critical point due to Griffith singularities, giving rise to a
divergence of the linear and nonlinear susceptibility already in the
disordered phase.  In a recent finite size scaling analysis of results
obtained via Monte Carlo simulations and the use of the
transfer-matrix formalism \cite{Crisanti94} discrepancies to this
scenario were found. This may well be a consequence of the smallness
of the lattice sizes studied ($L\le16$), since for this situation one
might expect {\it typical} rather than {\it average} results.

In two and three dimensions, which are most relevant for the above
mentioned experiments, no analytical results are known --- apart
from renormalization group calculations using the Migdal-Kadanoff
approximation \cite{Boechat94,Continentino94}. For this reason
extensive Monte Carlo simulations have been performed recently in
two dimensions by Rieger and Young \cite{Rieger94c} and in three
dimensions by Guo, Bhatt and Huse \cite{Guo94}. Usually the
investigation of quantum systems via Monte Carlo methods are
hampered by various deficiencies, the sign problem being the
most notorious one in this respect (see \cite{Linden92} for a review).
In studying the Ising spin glass in a transverse field however, one
can exploit the fact that it can be mapped exactly onto a classical
Ising model described by a {\it real} Hamiltonian.
Using the Suzuki-Trotter formula \cite{Suzuki76} one can easily
show that the ground state energy of the
$d$--dimensional quantum mechanical model (\ref{tisg}) is equal to the
free energy of a $(d+1)$--dimensional classical model, where the extra
dimension corresponds to imaginary time, i.e.
\be
\ba{ccl}
-{\textstyle  E( T = 0) \over \textstyle L^d} & = &
\lim_{T\to 0} {\textstyle T \over \textstyle L^d} {\rm Tr\;} e^{-\beta H}\\
 & & \\
 & = & {\textstyle 1 \over \textstyle \Delta\tau}
{\textstyle 1 \over \textstyle L_\tau L^d} {\rm Tr\;}  e^{-{\cal S}}
\ea
\label{suzuki}
\ee
where the imaginary time direction has been divided into $L_\tau$ time
slices of width $\Delta\tau$ ($\Delta\tau L_\tau = \beta$), and the
effective classical action, $\cal S$, is given by
\be
{\cal S} = -\sum_{\tau}\sum_{\langle ij\rangle}
K_{ij} S_{i}(\tau) S_{j}(\tau)
-\sum_{\tau}\sum_i K S_{i}(\tau) S_{i}(\tau+1)
\left(-\sum_{\tau}\sum_i H S_{i}(\tau)\right)\;,
\label{class}
\ee
where the $S_i(\tau)=\pm1$ are classical Ising spins, the indices $i$
and $j$ run over the sites of the original $d$--dimensional lattice
and $\tau = 1,2,\ldots,L_\tau$ denotes a time slice. In equation
(\ref{class}),
\be
\ba{rcl}
K_{ij} & = & \Delta\tau J_{ij} \\
H & = & \Delta\tau h\\
\exp(-2 K) & = & \tanh(\Delta\tau \Gamma)\\
\ea
\ee
One has the {\em same} random interactions in each time slice.
In order to fulfill the second equality in (\ref{suzuki}) precisely,
one has to perform the limit $\Delta\tau\to 0$, which implies
$K_{ij} \to 0$ and $K \to \infty$. However, the universal properties
of the phase transition are expected to be independent of $\Delta\tau$ so
we take $\Delta\tau = 1$ and set the standard deviation of the
$K_{ij}$ to equal $K$. Thus $K$, which physically sets the relative
strength of the transverse field and exchange terms in (\ref{tisg}), is
like an inverse ``temperature'' for the effective classical model in
(\ref{class}).

One sees that the (d+1)-dimensional classical model (\ref{class})
should order at low ``temperature'' (or coupling constant $K$) like a spin
glass in the $d$ spatial dimensions and ferromagnetically in
the imaginary time direction. From this one concludes
the existence of two different diverging length-scales in the classical
model (\ref{class}): one for the spatial (spin glass)-correlations,
which is $\xi$, and one for imaginary time (ferromagnetic) correlations,
which is $\xi_\tau$. Thus in the representation (\ref{class}) the link between
statics and dynamics in the original quantum model (\ref{tisg})
becomes most obvious. Correspondingly, to analyze the critical properties
of the extremely anisotropic classical model (\ref{class}) one has to
take into account these two length scales via anisotropic finite size scaling
\cite{Binder89}.

Monte Carlo simulations of the classical model (\ref{class})
are straightforward --- it turns out that sample-to-sample
fluctuations are significant, for which reason one has to
do an extensive disorder average \cite{Rieger94c,Guo94}.
However, the finite size scaling analysis is complicated by
the fact that due to the existence of two diverging length scales
$\xi$ and $\xi_\tau$ one has to deal with two independent
scaling variables: as usual $L/\xi$ and in addition the
shape (or aspect ratio) $L_\tau/L^z$ of the system \cite{Binder89}.
Thus, with the usual definition of a spin glass overlap
$Q=L^{-d}L_\tau^{-1}\sum_{i,\tau} S_i^a(\tau) S_i^b(\tau)$
for the classical system, the dimensionless combination of
moments of the order-parameter $g_{\rm av}$ obeys
\be
g_{\rm av}(K,L,L_\tau)=
0.5[3-\langle Q^4\rangle/\langle Q^2\rangle^2]_{\rm av}
\sim\tilde{g}_{\rm av}(\delta L^{1/\nu},L_\tau/L^z)\;.
\label{hypo}
\ee
In isotropic systems one has $z=1$, which makes the aspect ratio
constant to one for the choice $L=L_\tau$ and in order to
determine the critical coupling $K_c$ one exploits the fact that
$g_{\rm av}(K,L,L)$ becomes independent of $L$ for $K=K_c$ (see e.\
g.\ \cite{Bhatt87}). In the present case of a very anisotropic system
$z$ is not known a priori and one has to vary three different
system parameters to obtain an estimate for $K_c$ and $z$ (and other
exponents).  The following method \cite{Rieger94c,Guo94} enhances the
efficiency of such a search in a three-parameter space and also
produces reliable estimates for the quantities of interest: In the
limit $L_\tau\gg L^z$ the classical $(d+1)$-dimensional classical
system is quasi-one-dimensional, and in the limit $L_\tau\ll L^z$ the
system is quasi-$d$-dimensional and well above its
transition ``temperature'' in $d$ dimensions (which is even zero for
$d=2$). Therfore one has $\tilde{g}_{\rm av}(x,y)\to0$ for $y\to0$ and
for $y\to\infty$.  Hence, for fixed $x$, $\tilde{g}_{\rm av}(x,y)$
must have a maximum for some value $y=y_{\rm max}(x)$. The value of
this maximum decreases with increasing $L$ in the disordered phase $K
< K_c$ (where $\delta = (K_c / K - 1) > 0$) and increases with
increasing $L$ in the ordered phase. This criterion can be used to
estimate the critical coupling, as exemplified in figure
(\ref{fig_g}).  If one plots $g_{\rm av}(K_c,L,L_\tau)$ versus
$L_\tau/L^z$ with the correct choice for the dynamical exponent $z$,
one should obtain a data-collaps for all system sizes $L$. Finally one
uses systems with fixed aspect ratio $L_\tau / L^z$ to determine
critical exponents via the usual one--parameter finite size scaling.

Various scaling predictions can be made if one supposes a conventional second
order phase transition to occur at some critical ``temperature''
$K_c$ for the classical model. Hyperscaling in this particular situation
would imply \cite{Fisher89}
\be
2-\alpha=\nu(d+z)\;.\label{hyperz}
\ee
As usual one has $\gamma=\nu(2-\eta)$, where $\eta$ is defined via the decay
of correlations at criticality:
\be
\ba{ccccc}
C(r)& =    & [\langle S_i(\tau)S_{i+r}(\tau)\rangle^2]_{\rm av}
    & \sim & r^{-(d + z - 2 + \eta)}\;,\\
 & & & &\\
G(t)& =    & [\langle S_i(\tau)S_{i}(t+\tau)\rangle]_{\rm av}
    & \sim & r^{-(d + z - 2 + \eta)/2z}\;,
\ea
\ee
and from (\ref{hyperz}) one gets via $\alpha+2\beta+\gamma=2$ the
relation $2\beta/\nu=d+z-2+\eta$.
The uniform magnetic susceptibility defined via
\be
\chi_F = \frac{\textstyle
\partial [\langle\sigma_i^z\rangle]_{\rm av}}{\textstyle\partial h}
\sim |\delta|^{-\gamma_f}
\ee
with respect to the quantum mechanical Hamiltonian (\ref{tisg}),
is related to the integrated onsite correlation function
of the classical model (\ref{class}): $\chi_F\sim\sum_t G(t)$
and therefore $\gamma_f=\beta - \nu z$. Analogously the divergence of the
magnetic nonlinear susceptibility for the quantum system
\be
\chi_{nl} = \frac{\textstyle\partial^3[\langle\sigma_i^z\rangle]_{\rm av}}
{\textstyle\partial h^3}
\sim |\delta|^{-\gamma'}
\ee
can be estimated via the spin glass susceptibility
of the classical model \cite{Rieger94c,Guo94} giving
$\gamma'=\nu(2-\eta+2z)$.

In the following table we list the results obtained so far
in various dimensions
\be
\ba{|c||c|c|c|c|c|c|}
\hline
 & d=1 \cite{Fisher92} & d=1 \cite{Crisanti94} & d=2 \cite{Rieger94c} & d=3
\cite{Guo94} & d=3 \cite{Boechat94} & d\ge6 \cite{Ye93,Miller93}\\
\hline
z & \infty & \sim 1.7 & 1.50\pm0.05  & \sim 1.3 & \sim1.4 & 2 \\
\nu & 2  & \sim1 & 1.0\pm0.1 & \sim 0.8 & \sim0.87  & 1/4\\
\eta     & 0.38  & \sim0.4 & \sim0.5 & \sim0.9 & -  & 2\\
\gamma_f & div.\ & \sim2.3 & \sim0.5 & finite  & -  & finite\\
\gamma'  & div.\ & -       & \sim4.5 & \sim3.5 & -  & 0.5\\
\hline
\ea
\ee
The symbol $div$.\ in the second column means that first and higher
derivatives of the magnetization diverge already in the disordered
phase. The word $finite$ means that in three
dimensions and for $d$ larger than the upper critical dimension the
uniform susceptibility does not diverge at the critical point. The
results in the second column \cite{Fisher92} were obtained within a
renormalization group calculation, those in columns 3 to 5
\cite{Crisanti94,Rieger94c,Guo94} with Monte Carlo simulations, column
6 shows the result of a Migdal-Kadanoff RNG calculation (in
\cite{Continentino94} also $\gamma$ has been calculated in this way
giving $\gamma=1.2$, which is far off the MC-value) and the last column
depicts analytical results from mean-field theory
\cite{Ye93,Miller93}.

Note that the results reported so far are for the zero-temperature
critical behavior of the quantum model (\ref{tisg}). To make contact
with the recent experiments mentioned above, which are done in the
vicinity of the quantum critical point at {\it finite} temperature,
one can exploit the following scaling prediction. For finite temperatures the
scaling variable of e.g.\ the nonlinear susceptibility is $\xi_\tau\cdot T$
since a finite temperature for the quantum system (\ref{tisg}) implies
a finite length $L_\tau\sim T^{-1}$ in the imaginary time direction
of the classical model (\ref{class}). Hence one has
\be
\chi_{nl}(T,\delta)\sim\delta^{-\gamma'}\tilde{\chi}_{nl}\Bigl(
\frac{T}{\delta^{z\nu}}\Bigr)\;,
\ee
with $\tilde{\chi}_{nl}(x)\to{\rm const.\ }$ for $x\to0$ and
$\tilde{\chi}_{nl}(x)\to x^{-\gamma'/z\nu}$ for $x\to\infty$ (in
order to cancel the divergent prefactor at finite $T$ if
$\delta\to0$). Thus one has for $\Gamma=\Gamma_c$
\be
\chi_{nl}\sim T^{-\gamma'/z\nu}
\label{tsusc}
\ee
and analoguously e.g.\ $\chi_F\sim T^{-\gamma_F/z\nu}$.  Equation
(\ref{tsusc}) implies in three dimensions a strong divergence
of the nonlinear susceptibility (when approaching $T=0$)
with a power $\sim2.7$ that is very close to the one for classical
Ising spin glasses ($\sim2.9$ see section 2). This is in striking
contradiction to the observation made in the experiments mentioned in
the introduction \cite{Wu93} and in order to clarify this discrepancy
further work is necessary.

Another very important issue is the effect of Griffith singularities
on the properties of quantum spin glasses.  As already observed by
McCoy and Wu \cite{McCoy68,McCoy69b} for the one-dimensional quantum
model (corresponding to the 2-dimensional classical model with layered
disorder) their existence lead to divergencies in the magnetic
susceptibility already in the disordered phase near the critical point
(see also \cite{Shankar87,Nieuwenhuizen89b,Fisher92}). Thill and Huse
\cite{Thill94} recently presented a droplet theory for quantum spin
glasses, where they also predict algebraically decaying correlations
in this region. This implies that the effect of Griffith singularities
on the time-persistence of autocorrelations is much stronger for
quantum systems at zero temperature than their effect for classical
models at finite temperatures, where only an enhanced power law is
predicted \cite{Randeira85}.  Preliminary results of Monte Carlo
simulations of the two-dimensional transverse field Ising spin glass
at zero temperature give already strong support for this scenario to
be correct \cite{Rieger95}.

Finally we should mention that also the more general case of
spin-$\frac{1}{2}$ Heisenberg model with quenched disorder, described
by the Hamiltionian
\be
H=\sum_{\langle ij\rangle} J_{ij} \{
\sigma_i^x \sigma_j^x + \delta_{ij}(\sigma_i^y \sigma_j^y + \sigma_i^z
\sigma_j^z) \} -\sum_i \underline{h}\cdot\underline{\sigma}
\label{heisen}
\ee
is of great interest, in particular because e.g.\ the
copper-oxygen layers of high-T$_c$ superconductors are supposed to be
good physical realizations of such a model in two-dimensions
\cite{Sandvik94}. The additional disorder-parameters $\delta_{ij}$
introduce a random anisotropy without destroying the XY-symmetry and
the limit $\delta_{ij}\to0$ reproduces the Ising case considered
above. For vanishing external field $\underline{h}=0$ the strength of
the disorder $([J_{ij}^2]_{\rm av}-[J_{ij}]_{\rm av}^2)$ triggers the
zero-temperature quantum phase transition of the model (\ref{heisen}).
For the isotropic case $\delta_{ij}=1$ this scenario has been studied
in two dimensions by Sandvik and Vekic \cite{Sandvik94} very recently
with Monte Carlo simulations. Unfortunately, due to the notorious sign problem
occuring in (Quantum)-Monte Carlo simulations, only the non-frustrated
case ($J_{ij}>0$) has been considered. On the other side, much
progress could be made in the study of the XXZ-chain with quenched
disorder (i.e.\ model (\ref{heisen}) in one dimension
\cite{Doty92,Haas93,Runge94}), which is well suited for exact
diagonalization studies.

\section{Other models and conclusion}

Althought we have confined ourselves to {\it Ising} spin glasses
in this review, we finally would like to mention some of the other
spin glass like systems currently under investgation.
A concise review on this matter has already been given by
Young, Reger and Binder \cite{Young92}.

Many experimental spin glasses that most convincingly
show features of a spin glass transition at finite temperatures
are Heisenberg-like systems (e.g.\ CuMn and Eu$_{x}$Sr$_{1-x}$S)
described by the (classical) Hamiltonian
\be
H=\sum_{\langle ij\rangle} J_{ij} \underline{S}_i \underline{S}_j\;,
\ee
where $\underline{S}_i$ are 3-component vector spins. There seems to
be ample numerical evidence (see \cite{Matsubara91} for references) that this
model (with short range interactions) does {\it not} possess a finite
$T$ spin glass transition --- which is an unsatisfactory discrepancy
with the experimental situation (in four dimensions indications for a
finite-$T$ transition were found in a recent Monte Carlo study
\cite{Coluzzi94}). However, it has been demonstrated
by Matsubara et al.\ \cite{Matsubara91} via numerical simulations
using a hybrid Monte Carlo spin dynamics method
that an anisotropy term modelled by
\be
H_{\rm anis.}=\sum_{\langle ij\rangle}\sum_{\alpha\ne\beta}
D_{ij}^{\alpha\beta} S_i^\alpha S_j^\beta
\ee
induces a finite temperature phase transition (see also the
defect-wall renormalization-group study of Gingras \cite{Gingras93}),
which gives further support to the claim that all three-dimensional
spin glasses with anisotropy should have a phase transition in the
universality class of the short range Ising spin glass model \cite{Bray86}.

Another generalization of the Ising models considered here are so
called Potts glasses (see \cite{Young92} for a review), where each
spin can take on $p$ values instead of only $p=2$ in the Ising case.
It seems that in three dimensions for $p=3$ this model
shows a spin glass transition only at zero temperature
\cite{Scheucher90}. On the mean-field level it has been shown
\cite{Kirkpatrick87} that the p-state Potts glasses and spin glass
models with p-spin interactions show very similar static and dynamic
behavior for certain values of p. The question, whether the situation
is similar in finite dimensions, has been discussed in
\cite{Rieger92}, but due to the fact that on a finite dimensional
lattice the definition of a model with p-spin interactions is to some
extend arbitrary, it is not resolved yet. Furthermore it has been observed
\cite{Kirkpatrick87} that the dynamical saddle-point equations of both
mean-field spin glass models (the Potts model and the p-spin model)
have the same mathematical structure as that arising in mode-coupling
theory for the structural glass transition \cite{Jaeckle86}. This
latter similarity also occurs for orientational glasses,
(for a review see \cite{Binder92}), where the spins are replaced by
quadrupolar moments.

Very similar to the XY spin glass is the so-called gauge glass
defined by the Hamiltonian
\be
H=\sum_{\langle ij\rangle} J_{ij}\cos(\varphi_i-\varphi_j-A_{ij})\;,
\label{gauge}
\ee
where $\varphi\in[0,2\pi[$ are phase-variables and $A_{ij}\in[0,2\pi[$ is
a quenched random vector potential. This model has been introduced by
Shih et al.\ \cite{Shih84} to describe granular superconductors in an
external field. It has been discussed that the gauge glass (\ref{gauge})
might describe some of the low temperature physics of vortex-glasses
(see \cite{Blatter94}, chapter 7 for a review), which is the proposed
\cite{Fisher89a} "true" superconducting state of high-T$_c$ materials.
Of particular interest in this respect is the existence of a gauge glass
transition in three dimensions. Monte Carlo simulations
\cite{Huse90,Reger91,Reger93} and defect-wall energy analysis \cite{Gingras92}
of the gauge glass found indeed a phase transition in three dimensions
at finite temperatures. Recently, however, it was demonstrated
\cite{Bokil94} that screening effects might destroy this transition.
Moreover, it is not clear whether the gauge glass and the vortex glass
really belong to the same universality class (\cite{Blatter94},
chapter 7).  The latter has gained much interest for the analytically
tractable case in 1+1 dimensions, which has been studied numerically
via Monte Carlo simulations only very recently \cite{Batrouni94,Cule94}.
Descrepancies to existing theories rekindled the interest in this
model that still seems not to be fully understood.

Finally we would like to mention the numerical investigation of a
Hubbard model of interacting bosons within a quenched random potential
\cite{Fisher89}. Similar to the case of the quantum spin glass discussed
in the last section this quantum mechanical Hamiltonian can be mapped
onto a real classical Hamiltonian in d+1 dimensions. Monte Carlo
simulations of this model in two dimensions \cite{Sorensen94,Wallin94}
showed the existence of a transition from a super conducting phase to
an insulating Bose glass phase.

In summary we have seen that the application and usefulness of Monte
Carlo simulations of strongly disordered systems has spread from Ising
spin glasses and random field systems to very diverse fields including
structural glasses, superconductors and "dirty bosons".  However, also
the most "simple" spin glass model in finite dimensions, the EA-model
with Ising spins, gains again increasing interest. Apart from many
fundamental questions still being unsolved, like the existence of a
transition within an external field, recent years' research activities
began on one side to focus on quantum effects in spin glasses and on
the other side to shift from critical behavior to non-equilibrium
dynamics.  In both cases the main stimulus came from experiments and
Monte Carlo simulations play an important role in learning to
understand the newly observed phenomena. Thus, to conclude, spin
glasses are far from being dead --- they are still aging.

\vskip0.8cm
\noindent{\large \bf Acknowledgement}
\vskip0.3cm

During recent years I had the opportunity to discuss with many experts on
this field. In this respect I would like to express my gratitude
to J.\ Adler, D.\ Belanger, J.\ P.\ Bouchaud, D.\ Chowdhury, A.\
Crisanti, S.\ Franz, H.\ O.\ Heuer, H.\ Horner, D.\ Huse, W.\ Kinzel,
W.\ Kleemann, L.\ Mikheev, T.\ Nieuwenhuizen, P.\ Nordblad, M.\
Schreckenberg, P.\ Sibani, D.\ Stauffer, T.\ Nattermann, E.\ Vincent
and A.\ P.\ Young. My special thanks also to J.\ Kisker and
B.\ Steckemetz. This work work was performed within the SFB 341
K\"oln--Aachen--J\"ulich, supported by the DFG.

\vfill
\eject



\newpage

\noindent{\large \bf Figure Captions}\\[1ex]

\begin{enumerate}

\item Contours of $\chi_{SG}=1$ (obtained via Monte Carlo
simulations) for two-, three- and four-dimensional spin glasses in an
external field compared to series expansions \cite{Singh91}.
 From ref.\ \cite{Grannan91}. \label{fig_cont}

\item Waiting time experiments: Relaxation of the thermo-remanent
magnetization for different waiting times $t_w$. The procedure mentioned in
the text is sketched in the inset: the sample (CdCr$_{1.7}$In$_{0.3}$S$_4$)
is cooled from above $T_g=16.7K$ down to $0.6T_g=10K$ in a 15 Oe field,
kept in the field during $t_w$, then the field is cut off.
 From ref.\ \cite{Vincent92}. \label{fig_age}.

\item  Top: Decay of the remanent magnetization of
(Fe$_x$Ni$_{(1-x)}$)$_{75}$P$_{16}$B$_6$Al$_3$ as function of time $t$
for various temperatures ($T_g=22.6K$) in a log-log plot.
Bottom: Exponent $m$ as a function of temperature obtained from a fit
of the data of the upper figure to an algebraic decay
$M_S(t)\propto t^{-m(T)}$ for $T\le0.97 T_g$. From ref.\
\cite{Granberg87}. \label{fig_rem}

\item The autocorrelation function $C(t,t_w)$ for the three-dimensional
EA model with Gaussian couplings at $T=0.5$ ($T_g\approx0.9$ \cite{Bhatt87}).
The system size is $L=24$ and the data are averaged over 128 samples.
 From ref.\ \cite{Kisker94b}, see \cite{Rieger93b} for similar data in case
of the $\pm1$-distribution. \label{fig_ctw}

\item The correlation length $\xi(t_w)$ of the two-dimensional EA-model
with Gaussian couplings as a function of the waiting
time $t_w$ in a log-log plot. The straight lines are least-square fits toan
algebraic growth law (\ref{alg}) with exponents $\alpha(T)=0.044$, $0.066$
and $0.081$ for temperatures $T=0.2$, $0.3$ and $0.4$, respectively.
Data taken from ref.\ \cite{Rieger94b}. \label{fig_alg}.

\item Phase diagram of the diluted antiferromagnet in a uniform field:\\
$H=\sum_{\langle ij\rangle} \epsilon_i\epsilon_j S_i S_j
- B\sum_i\epsilon_i S_i$ with 50\% dilution ($\epsilon=0$ or $1$ with
probability $1/2$). PM means paramagnetic phase, AFM means antiferromagnetic
long range order and SG is the intermediate spin glass phase
(as explained in the text). From ref.\ \cite{Nowak91}. \label{fig_phase}.

\item Phase diagram of the diluted dipolar coupled Ising spin glass
LiHo$_{0.167}$Y$_{0.833}$F$_4$ in the $\Gamma$--$T$ plane. Filled circle
follow from the dynamical behavior of the linear susceptibility \cite{Wu91},
open circles from measurements of the nonlinear susceptibility. We are
interested in the region close to $T=0$, $\Gamma=1$. From ref.\ \cite{Wu93}.
\label{fig_qsg}

\item The dimensionless cumulant $g_{\rm av}$ for the (2+1)-dimensional
classical model (\ref{class}) for three different values of the
coupling constant versus the system size $L_\tau$ in the imaginary
time direction. The system size in the space direction is
$L=4$ ($\diamond$), $6$ ($+$), $8$ ($\Box$), $12$ ($\times$) and
$16$ ($\triangle$). Since the maximum of $g_{\rm av}(L_\tau)$ is
roughly independent of $L$ at $K^{-1}\approx3.30$ one concludes that the
latter value is the critical coupling constant. From ref.\ \cite{Rieger94c}.
\label{fig_g}.

\end{enumerate}

\end{document}